\documentclass[aps,10pt,prd,twocolumn,floatfix,amsmath,amssymb,nofootinbib,superscriptaddress]{revtex4-2}
\usepackage[utf8]{inputenc}
\usepackage{graphicx,bm,bbm,color,amssymb,amsmath,float,caption}
\usepackage{slashed,verbatim,ulem}
\usepackage{xcolor, soul}
\usepackage{appendix}
\usepackage{float}
\usepackage{comment}
\usepackage{subcaption}
\usepackage{hyperref}
\hypersetup{colorlinks,linkcolor={blue},citecolor={magenta}}

\newcommand{\bea}{\begin{eqnarray}}
\newcommand{\eea}{\end{eqnarray}}

\def\nn {\nonumber}
\newcommand{\bl}{\color{blue}}
\newcommand{\rd}{\color{red}}

\begin{document}

\title{Non-perturbative heavy quark diffusion coefficients in a weakly magnetized thermal QCD medium}

\author{Debarshi Dey}
\email{debarshi@iitb.ac.in}
\affiliation{Indian Institute of Technology Bombay, Mumbai 400076, India}

\author{Aritra Bandyopadhyay}
\email{aritra.bandyopadhyay@e-uvt.ro}
\affiliation{Department of Physics, West University of Timişoara, Bd. Vasile Pârvan 4, Timişoara 300223, Romania}

\author{Santosh K. Das}
\email{santosh@iitgoa.ac.in}
\affiliation{School of Physical Sciences, Indian Institute of Technology Goa, Ponda-403401, Goa, India}

\author{Sadhana Dash}
\email{sadhana@phy.iitb.ac.in}
\affiliation{Indian Institute of Technology Bombay, Mumbai 400076, India}
	
\author{Vinod Chandra}
\email{vchandra@iitgn.ac.in}
\affiliation{Indian Institute of Technology Gandhinagar, Gandhinagar 382055, India}

\author{Basanta K. Nandi}
\email{basanta@iitb.ac.in}
\affiliation{Indian Institute of Technology Bombay, Mumbai 400076, India}

\begin{abstract}
In this work, the perturbative and non-perturbative contributions to the heavy quark (HQ) momentum 
($\kappa$) as well as spatial ($D_s$) diffusion coefficients are computed in a weak background magnetic field. The formalism adopted here involves calculation of the in-medium potential of the HQ in a weak magnetic field, which then serves as a proxy for the resummed gluon propagator in the calculation of HQ self-energy ($\Sigma$). The self-energy determines the scattering rate of HQs with light thermal partons, which is subsequently used to evaluate $\kappa$ and $D_s$. It is observed that non-perturbative effects play a dominant role at low temperature. The spatial diffusion coefficient $2\pi T D_s$, exhibits good agreement with recent LQCD results. These findings can be applied to calculate the heavy quark directed flow at RHIC and LHC energies. An extension of this formalism to the case of finite HQ momentum has also been attempted.
\end{abstract}

\maketitle

\section{Introduction}
 Heavy quarks (HQs)~\cite{Rapp:2018qla,Das:2406.2024}, such as charm and bottom quarks, have been extensively studied in the heavy-ion community as valuable probes for investigating the properties of hot and dense quark matter created in relativistic heavy-ion collisions (HIC).
Due to their large masses, HQs are produced primarily through hard scatterings in the early stages of these collisions.  
Also, because $M_Q\gg T$, where $M_Q$ is the heavy-quark mass, and $T$ is the temperature of the thermal QCD medium, they do not fully thermalize with the medium. Consequently, HQs act as effective tracers of the entire space-time evolution of the quark matter until freeze-out
~\cite{Dong:2019unq,QGP4,He:PPNP130'2023}. After freeze-out, they can be reliably tracked as they hadronize into various $D$ (charm) and $B$ (bottom) mesons. 
Because they are too massive to be regenerated through thermal scatterings, their yields are essentially fixed early on, leading to a well-defined initial momentum distribution. These characteristics collectively establish HQ's as excellent probes of the hot and dense quark matter. 

As the HQ traverses the medium, it interacts with the thermal quarks and gluons. This leads to the HQ losing energy and its momentum getting diffused, characterized by the heavy quark transport coefficients, \textit{viz.,} drag and momentum diffusion coefficients~\cite{Svetitsky:PRD37'1988,Mustafa:PRC72'2005}.  
 Due to their substantial mass, HQ dynamics can be described using the equations for a Brownian particle, more specifically - the Langevin or Fokker-Planck equation which treats the heavy quarks as external particle experiencing random kicks from thermal partons (light quark and gluons) within the medium. 
 The HQ drag and diffusion coefficients play crucial role in shaping heavy quark phenomenology and, consequently, theoretical predictions related to relevant experimental observables. For instance, the HQ energy loss is expected to show up in experimental observables such as the $R_{AA}$ and $v_2$. 
 Contrary to initial expectations~\cite{Armesto:PLB637'2006}, the HQ nuclear modification factor ($R_{AA}$)~\cite{PHENIX:2005nhb} at intermediate $p_T$ was observed to be small (large suppression), and the elliptic flow ($v_2$) ~\cite{PHENIX:2006iih} was almost as large as that of light hadrons. Phenomenologically, the simultaneous description of the HQ $R_{AA}$ and $v_2$ remained a challenge. A step towards its resolution was put forward in ~\cite{Das:PLB747'2015}, where the authors employed $T$-dependent coupling constant for their analysis.

 Leading order pQCD analysis of the HQ dynamics in a thermal medium either by solving the Fokker-Planck equation or via evaluating the HQ in-medium self energy, has been a popular and successful approach~\cite{Braaten:PRD44'1991,Mustafa:PRC57'1998,Mustafa:PRC72'2005,Das:PRC80'2009,Das:PRC90'2014,Moore:PRC71'2005,Thoma:NPB351'1991,Braaten:PRD44'1991,Monteno:JPG38'2011,He:PLB735'2014,Beraudo:EPJC75'2015,Das:PLB747'2015,Sadofyev:PRD93'2016,Akamatsu:PRC92'2015,Kurian:PRD100'2019,Ruggieri:2022kxv,Gossiaux:2008jv,Song:2015sfa,Cao:PRC84'2011}. HQ drag and diffusion has been studied in strongly coupled plasma as well, using the AdS/CFT correspondence~\cite{Rajagopal:JHEP10'2015,Jorge:PRD74'2006,Solana:PRD74'2006}. In ~\cite{Simon:PRL100'2008}, a perturbative 
Next-to-leading-order (NLO) calculation of the HQ diffusion was carried out for the first time, in which the authors showed that the NLO correction was very large, which implied poor perturbative convergence. Non-perturbative studies of HQ dynamics then began primarily via Lattice QCD (LQCD) simulations. Initial  LQCD computations of the HQ diffusion coefficient involved evaluating the momentum diffusion coefficient $\kappa$ up to $\mathcal{O}(T/M_Q)$ , from color electric field correlators connected by Wilson lines~\cite{Banerjee:PRD85'2012,Simon:JHEP'2009,Francis:PRD92'2015,Brambilla:PRD102'2020,Altenkort:PRD103'2021}. Recently, advances have been made wherein the color magnetic contribution (which is suppressed by a factor of $1/M_Q$ relative to the color-electric correlator) is also taken into account~\cite{Bouttefeux:JHEP'2020,Banerjee:JHEP'2022,Altenkort:PRL132'2024,Altenkort:PRD109'2024}. 

Not only has the effect of external electromagnetic fields on the HQ dynamics been investigated phenomenologically in the literature, HQ dynamics has also been proposed as a probe of those fields. For instance, in ~\cite{DAS:PLB768'2017},
the authors propose the directed
flow ($v_1$) of HQ's to be a good probe of the magnetic field
generated in noncentral heavy-ion collisions.  The heavy quark $v_1$ is also computed for the small system~\cite{Sun:2023adv} produced in the pPb collisions in the presence of electromagnetic field.
In ~\cite{Chatterjee:PLB798'2019}, the combined effect of the initial tilt of the QGP
fireball and large EM fields on the HQ $v_1$ is studied. Experimental measurements at both RHIC~\cite{STAR:2019clv} and LHC~\cite{ALICE:2019sgg} have shown that the directed flow of the D-meson is non-zero. To compute the HQ directed flow in presence of the electromagnetic fields, it is essential to consider the influence of the electromagnetic field on HQ transport coefficients along with the bulk evolution. The nuclear suppression factor and elliptic flow of heavy mesons are other experimentally measured observables that can be influenced by the anisotropic HQ transport coefficients due to the presence of an electromagnetic field. Several pQCD studies of heavy quark diffusion in the presence of magnetic fields have also been carried out~\cite{Singh:arxiv'2020,Singh:JHEP5'2020,Fukushima:PRD93'2016,Bandyopadhyay:PRD105'2022,Dey:PRD109'2024,Bandyopadhyay:2023hiv,Satapathy:PRC109'2024}. The effect of momentum
anisotropy on the dynamics of HQ has also been studied
recently~\cite{Avdhesh:PRC105'2022,Jai:PRD108'2023}. Apart from the thermalized QGP, the heavy quark diffusion coefficient in the initial pre-equilibrium phase has also been explored in~\cite{Ruggieri:2018rzi, Liu:PRD103'2021, Stanislaw:EPJA54'2018,Pandey:2023dzz,Boguslavski:2020tqz,Khowal:2021zoo}.

It turned out that the prediction of the HQ spatial diffusion coefficient ($D_s$) from LQCD studies was an order of magnitude less than LO pQCD studies~\cite{He:PPNP130'2023}. This, coupled with recent lattice studies incorporating color-magnetic interactions into their analysis, motivated us to explore non-perturbative HQ diffusion in the presence of a weak magnetic field. Analytically, one of the ways of incorporating non-perturbative effects in the HQ dynamics is via the HQ potential, although to do this, one has to consider the static limit of the HQ. Once the potential is evaluated, it can be used as a proxy for the gauge boson propagator within the static limit of the HQ, to calculate the scattering matrix elements of the HQ with the light thermal partons of the medium~\cite{Hees:PRL100'2008,Riek:PRC82'2010,Xing:PLB838'2023}. The heavy quark potential in the presence of magnetic field has been evaluated in the past~\cite{Lata:PRD97'2018,Hasan:PRD102'2020,Indrani:EPJC83'2023}. Using the HQ potential to evaluate the HQ drag and diffusion coefficients in a strong background magnetic field has been attempted recently~\cite{mazumder2023:arxiv}. In this paper, we follow a similar procedure, wherein we calculate the heavy-quark potential in the presence of a weak magnetic field. In particular, the information about the magnetic field enters through the HTL resummed gauge boson (gluon) propagator, which is then used to calculate the potential. The non-perturbative effects are incorporated via a non-perturbative ansatz for the gluon propagator which gives rise to a string-like confining term in the coordinate space, \textit{i.e.}, the gluon propagator is assumed to be a sum of the usual perturbative (coulomb-like) term and a non-perturbative term~\cite{Guo:PRD100'2019}. We evaluate the HQ momentum diffusion coefficents ($\kappa$) both in the static limit and beyond, and also the scaled HQ spatial diffusion coefficient ($2\pi TD_s$). Comparison of the latter with lattice results is also presented.

The paper is organized as follows: In section \ref{sec2}, the formalism of the work is outlined, and all relevant formulae are presented. Then in section \ref{sec3}, the calculation of the heavy quark potential is shown. Using the HQ potential, the scattering rate $\Gamma$ is evaluated in section \ref{sec4}. In section \ref{sec5}, the momentum diffusion coefficients are presented. The results are discussed in section \ref{sec6}. Finally, the summary and conclusion are presented in section \ref{sec7}.


\section{Formalism}\label{sec2}
The typical scale of momentum transfer to the HQ from the light thermal partons is $T$, where $T$ is the temperature of the QGP. Since the HQ thermal momentum [$\mathcal{O}(\sqrt{M_QT})$] is much larger than this typical exchanged momentum, the scatterings can be treated as random thermal kicks, which accumulate over time to evolve the HQ momentum. This essentially is the motivation to apply the Langevin equation of motion to analyse the HQ momentum evolution in a thermal medium~\cite{Reif2009,Moore:PRC71'2005} 
	\begin{equation} \label{reif}
		\frac{dp_i}{dt}=\xi_i(t)-\eta_Dp_i\,, \qquad \langle \xi_i(t)\xi_j(t')\rangle=\kappa\,\delta_{ij}\delta(t-t'),
	\end{equation}
where, $(i,j)=(x,y,z)$, $\xi$ is a random force which encodes the interaction of the HQ with the other degrees of freedom in the thermal bath (light quarks and gluons), $\eta_Dp$ is the friction term that works towards bringing the HQ momentum back to equilibrium, with $\eta_D$ being the drag coefficient.  $\xi$ is a rapidly fluctuating white noise force, whose auto-correlation is determined by the HQ momentum diffusion coefficient $\kappa$.  The rotational symmetry evident from Eq.~\eqref{reif} reflects the fact that the HQ has been considered to be static (\textit{i.e.}, $p\sim 0$) because of its large mass ($M_Q\gg T$). Hence we call $\eta_D$ and $\kappa$, respectively, the HQ drag and momentum diffusion coefficients within the static limit of HQ where they are related via the fluctuation dissipation relation:
\begin{equation}\label{fdt}
     \eta_D=\frac{\kappa}{2M_QT}.
\end{equation}
This subsequently leads to another phenomenologically important quantity called the HQ spatial diffusion coefficient $D_s$, defined as~\cite{He:PPNP130'2023}
\begin{equation}\label{Ds}
		D_s=\frac{T}{\eta_D M_Q},
	\end{equation}
which after using Eq.\eqref{fdt} becomes
\begin{equation}\label{ds}
    D_s=\frac{2T^2}{\kappa}.
\end{equation}  
    
 The presence of a background magnetic field breaks rotational symmetry, and thus, we have
    \begin{align}
	\frac{dp_{z}}{dt}&=\xi_z-(\eta_D)_{\parallel}p_z\,,\quad \langle \xi_z(t)\xi_z(t')\rangle=\kappa_{\parallel}(\bm{p})\,\delta(t-t')\label{lel}\\
	\frac{d\bm{p}_{\perp}}{dt}&=\bm{\xi}_{\perp}-(\eta_D)_{\perp}\bm{p}_{\perp}\,,\quad \langle \bm{\xi}_{\perp}(t)\bm{\xi}_{\perp}(t')\rangle=\bm{\kappa}_{\perp}(\bm{p})\,\delta(t-t')\label{let},
\end{align}
where the magnetic field is assumed to point along $\hat{z}$, so that $p_z$ is the longitudinal HQ momentum with $\kappa_{\parallel}$ and $(\eta_D)_{\parallel}$ being the longitudinal diffusion and drag coefficients. The remaining two directions ($\hat{x}$, $\hat{y}$) constitute the perpendicular direction. In addition to the magnetic field, if we also go beyond the HQ static limit and consider a finite momentum or velocity ($\bm{p} /\bm{v}$) of the HQ , that also defines a special direction, and thus, creates another anisotropy. In this case, the longitudinal ($\kappa_L$) and transverse ($\kappa_T$) momentum diffusion coefficients are defined with respect to the HQ velocity (momentum) $\bm{v}$ ($\bm{p}$). The interplay between these two anisotropy directions gives rise to the different drag and diffusion coefficients. In the simplistic case of $\bm{v}\parallel \bm{B}$, $(\parallel,\perp)$ is interchangeable with ($L,T$) in Eqs.(\ref{lel}, \ref{let}).

\begin{figure}[H]
    \centering
    \includegraphics[width=0.45\linewidth]{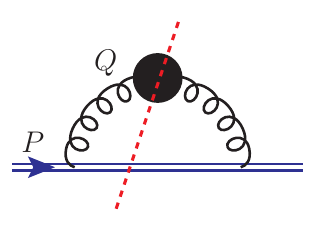}
     \captionsetup{justification=raggedright, singlelinecheck=false, format=hang, labelsep=period}
    \caption{Heavy quark one-loop effective self energy with resummed gluon propagator.}
    \label{figSE}
\end{figure}
\begin{figure}[H]
		\centering
		\includegraphics[width=0.95\linewidth]{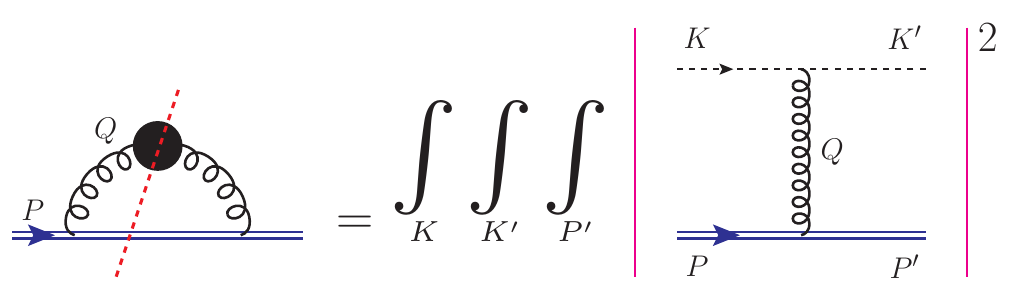}
         \captionsetup{justification=raggedright, singlelinecheck=false, format=hang, labelsep=period}
        \caption{Imaginary or cut part of the HQ self energy corresponds to $t$-channel scattering matrix element squared of the HQ with thermal quarks and gluons.}
        \label{resum}
	\end{figure}

While computing the HQ transport coefficients like drag or diffusion, the key quantity is the HQ scattering rate $\Gamma$. $t$- channel scattering of heavy quarks with thermal quarks and gluons are considered here in the evaluation of $\Gamma$. Gluon bremsstrahlung and Compton scattering processes are neglected since the former contributes only at higher order in $\alpha_s$, and the contribution of the latter is suppressed by powers of $M_Q/T$. The scattering rate $\Gamma$ can then be calculated from the imaginary part of the one-loop effective HQ self-energy, as given by Weldon's formula~\cite{Weldon:PRD28'1983}
	\begin{equation}\label{gamma}
	\Gamma(P , \bm{v})=-\frac{1}{2 E} [1-n_F(E)] \operatorname{Tr}\left[(\slashed{P}+M_Q) \operatorname{Im} \Sigma\left(p_0+i \epsilon, \bm{p}\right)\right],
    \end{equation}
where $\Sigma(p_0,\bm{p})$ is the one-loop effective HQ self-energy, shown in Fig.~\ref{figSE}. The blob in the gluon propagator of Fig.~\ref{figSE} represents the resummed gluon propagator.  In the kinematic region, where the exchanged gluon momentum is soft, $\Gamma$ turns out to be quadratically infrared divergent, thus necessitating resummation of the gluon propagator. Hard thermal loop (HTL) corrections to the gluon propagator contribute at leading order in the strong coupling $g$ and, therefore, have to be resummed, resulting in softening of the divergence in $\Gamma$ to only a logarithmic one~\cite{Braaten:PRD44'1991}. The cut (or imaginary) part of the diagram in Fig.~\ref{figSE} is proportional to $t$-channel scattering matrix element squared of the HQ with thermal quarks and gluons, as clearly indicated in Fig.~\ref{resum}. As such, this diagram describes the scattering of the HQ with both light quarks and gluons of the medium with soft ($\mathcal{O}[gT]$) $t$-channel gluon exchange.

Using Feynman rules, the self energy in Fig.~\ref{figSE} can be expressed as
\begin{equation}
\Sigma(P)=i g^2 \int \frac{d^4 Q}{(2 \pi)^4} \mathcal{D}^{\mu \nu}(Q) \gamma_\mu S(P-Q) \gamma_\nu\label{self_energy}
\end{equation}
The information about the magnetic field is encoded in the effective gauge boson propagator within a hot magnetised medium. In Eq.\eqref{self_energy}, $D^{\mu\nu}$ is the HTL resummed gluon propagator in the presence of a weak magnetic field~\cite{Karmakar:EPJC79'2019}, and $S(P-Q\equiv K)$ is the bare heavy quark propagator represented by the double lines in Figs.~\ref{figSE},\ref{resum}. In the present work, we have considered the magnetic field to be reasonably weak, \textit{i.e.}, we work in the regime $\frac{qB}{M_Q}\ll 1$, by the virtue of which we can ignore Landau quantization of the heavy quark energy levels~\cite{Singh:JHEP5'2020}, as the HQ mass scale is much larger than the scale defined by the difference in consecutive Landau levels:
\begin{equation}
    iS(P-Q\equiv K)=i\frac{\slashed{K}+M_Q}{K^2-M_Q^2}
\end{equation}

In QED, the bare photon propagator is given by $G^{\mu\nu}=g^{\mu\nu}/q^2$, where $g^{\mu\nu}$ is the metric, and $q$ is the photon momentum. The temporal component of the propagator is then $G^{00}=1/q^2$, whose Fourier transform is $1/(4\pi r)$. Thus, $e^2G^{00}$ gives the Coulomb potential. Extending the equivalence in this potential approach, we use the HQ in-medium potential $V(q)$ as a proxy for the resummed gluon propagator $D^{\mu\nu}$. In particular, the replacement in Eq.\eqref{self_energy} is $g^2D^{\mu\nu}\rightarrow V(q)$, which is possible when we consider the HQ to be approximately at rest in the rest frame of the plasma, \textit{i.e.}, $P^\mu = (M_Q,\bm{p}\sim 0)$ and only the temporal component of $D^{\mu\nu}$, \textit{i.e.}, $D^{00}$ contributes. 
The use of such a potential approach allows us to take into account the non-perturbative interactions via the non-perturbative part of the HQ potential. The evaluation of both the perturbative and non-perturbative parts of the HQ potential in the presence of a weak magnetic field in a hot QCD medium is discussed in Sec.\ref{sec3}.

The HQ self-energy thus calculated using the HQ potential leads us to the scattering rate $\Gamma$, Eq.\eqref{gamma}. From $\Gamma$, one can calculate the momentum diffusion coefficients as
 \begin{align}\label{kappa}
\kappa_{i}=\int d^3 q \frac{d \Gamma(E,v)}{d^3 q} q_{i}^2,
\end{align}
where $i=x,y,z$. $\frac{d \Gamma(E,v)}{d^3 q}$ is the scattering rate of the heavy quark via one-gluon exchange with thermal particles per unit volume of momentum transfer $\bm{q}$. It is interpreted as the differential probability per unit time for the heavy quark momentum to change by $\bm{q}$. From $\kappa_{i}$, one then obtains the corresponding drag coefficients via the fluctuation dissipation relation.

 \section{MICROSCOPIC DESCRIPTION OF THE HEAVY QUARK POTENTIAL}\label{sec3}

The heavy quark potential in vacuum is considered to be of the Cornell type with a Coulomb-like term and a string term as follows~\cite{Eichten:PRD17'1978}
 \begin{equation}\label{vac_pot}
     V_0(r)=-\frac{4}{3}\frac{\alpha_s}{r}+\sigma r,
 \end{equation}
 where $r$ is the distance between the heavy quark-antiquark pair, and $\sigma$ is the string tension, with a dimension of mass squared. 
In the presence of a thermal medium ($T\neq 0$, $B=0$), the HQ potential can be written  as~\cite{Liu:PRC97'2018}
 \begin{equation}
     V(r)=V_Y(r)+V_S(r)=-\frac{4}{3}\alpha_s\frac{e^{-m_Dr}}{r}-\sigma\frac{e^{-m_sr}}{m_s},
 \end{equation}
The first term represents a short-range Yukawa-type potential, while the second is a long-range confining term, making it non-perturbative. This generalized in-medium Cornell potential correctly reproduces the vacuum result while incorporating medium-induced screening of both Coulomb and confining interactions through temperature dependent screening masses $m_D$ and $m_s$. However, it should be remembered that the screening in both the terms originates from the interactions with the same medium partons, and thus, $m_D$ and $m_s$ are not independent. Hence, in the present study, we do not define two separate screening masses as above, when we work in the momentum space. Instead, the medium effects are incorporated into the vacuum perturbative HQ potential in momentum space $V_0^p(q)$ via a complex dielectric function $\epsilon(q)$. The thermal medium is immersed in a weak, constant magnetic field which influences the in-medium heavy quark potential. This potential is derived from the temporal part of the gauge boson propagator $D^{00}$ via $\epsilon(q)$,
\begin{equation}\label{diel}
    \frac{1}{\epsilon(q)}=-q^2D^{00}(q_0=0,q)
\end{equation}
The real and imaginary parts of $\epsilon(q)$ are then simply given as
\begin{align}
     \text{Re/Im}\frac{1}{\epsilon(q)}&=-q^2\,\text{Re/Im}D^{00}(q_0=0,q).\label{imeps}
\end{align}
This permittivity $\epsilon(q)$ contains all information about the medium. Thereafter, the in-medium HQ potential in momentum space is obtained by modifying the vacuum perturbative potential in momentum space $V_0^p$ by the dielectric function $\epsilon(q)$, \textit{i.e.},
\begin{equation}\label{pot}
     V(q)=V_0^p(q)\, \frac{1}{\epsilon(q)} ; ~~  V_0^p(q)=-\frac{4}{3}4\pi \alpha_s\frac{1}{q^2},
 \end{equation}
which is the Fourier transform of the first term in the R.H.S. of Eq.\eqref{vac_pot}. The inclusion of the confining term is achieved by phenomenologically adding a non-perturbative part in the perturbative gluon propagator~\cite{Eugenio:JHEP1'2006,Megias:PRD75'2007,Riek:PRC82'2010}. 
\begin{equation}
    D^{00}(q)=D_{\text{p}}^{00}(q)+D_{\text{np}}^{00}(q).
\end{equation}
    This non-perturbative term is chosen in a way that it reproduces correctly the confining term in vacuum. The perturbative and non-perturbative contributions in the propagator are thus independent. This will then lead to perturbative and non-perturbative contributions of $\epsilon(q)$ via Eq.\eqref{diel}, thus leading to a similar decomposition of $V(q)$ via Eq.\eqref{pot}:
    \begin{equation}\label{pot_decomp}
      V_Y(q)=V_0^p(q)\, \frac{1}{\epsilon_Y(q)}\,,\quad  V_S(q)=V_0^p(q)\, \frac{1}{\epsilon_S(q)},   
    \end{equation}
    with 
    \begin{align}
         \frac{1}{\epsilon_Y(q)}&=-q^2D^{00}_{\text{p}}(q_0=0,q)\label{epsy}\\  \frac{1}{\epsilon_S(q)}&=-q^2D^{00}_{\text{np}}(q_0=0,q).\label{epss}
    \end{align}
 As will be seen in section \ref{sec4}, it is the imaginary part of the HQ potential that is required for the calculation of the scattering rate, and subsequently, the momentum diffusion coefficients. So, we outline the derivation of the imaginary parts
of both the perturbative and non-perturbative parts of the propagator in the following subsections.

\subsection{Perturbative (Yukawa) part}
It will be seen in Sec. \ref{sec4} that the imaginary parts of the in-medium potentials will be used in the calculation of the scattering rate. $\text{Im}V(q)$ is obtained from $\text{Im}D^{00}$. We, therefore, present here a brief sketch of the derivation of $\text{Im}D^{00}$, evaluated in the real time formalism  ~\cite{Guo:PRD100'2019}. Imaginary part of the propagator (the 1-1 component) comes from the symmetric/Feynman propagator in Keldysh representation. 
\begin{equation}\label{imd11}
    \text{Im}D_{11}^L(q_0=0,q)=\frac{1}{2}D_F^L(q_0=0,q),
\end{equation}
where $L$ refers to the longitudinal component, $i.e.,$ $D^L\equiv D^{00}$. Now,
\begin{equation}\label{df}
    D_F^L(q_0,q)=[1+2n_B(q_0)]\text{sgn}(q_0)\left[D_R^L(q_0,q)-D_A^L(q_0,q)\right]
\end{equation}

Here, $R$, $A$, and $F$ stand for the retarded, advanced and the Feynman propagator, respectively. In the limit $q_0\to 0$, the Bose distribution function $n_B(q_0)$ can be expanded in powers of $q_0/T$ such that
\begin{equation}\label{df_exp}
       1+2n_B(q_0)=\frac{2T}{q_0}+\mathcal{O}(q_0^0)+\mathcal{O}(q_0)+\cdots  
\end{equation}
The Dyson-Schwinger equation allows us to calculate the perturbative resummed propagator from the bare propagator and the particle self-energy: 
 \begin{equation}\label{sde}
     D_{R/A}^{L,\,\text{p}}(q_0,q)=\left[q^2-\Pi_{R/A}^{L,\,\text{p}}(q_0,q)\right]^{-1},
 \end{equation}
where $q^2$ is the inverse of the bare gluon propagator, $\Pi^L\equiv \Pi^{00}$, and the superscript $\text{p}$ stands for perturbative. The gluon self energy  $\Pi_{R/A}^L(q_0,q)$ is computed using the Feynman diagram shown in Fig.~\ref{figSE} in the presence of a weak  magnetic field. This has already been done in the literature~\cite{Hasan:PRD102'2020,Karmakar:EPJC79'2019}. Thus, the steps involved in evaluating $\text{Im}D^{00}$ can be depicted as
  \begin{equation}
    \Pi_{R/A}^L\longrightarrow  D_{R/A}^L\longrightarrow D_F^L\longrightarrow \text{Im}D_{11}^L\equiv \text{Im}D^{00}.
 \end{equation}
 The details of the calculation of $\text{Im}D^{00}$ is provided in Appendix \ref{AppendixA}. The final expression reads
 \begin{widetext}
 \begin{align}\label{imdp}
\text{Im}D^{00}_{\text{p}}(q_0=0,q)=\frac{-\pi T\left[m_D^2-\sum_f\frac{g^2(q_fB)^2}{2\pi^2}\left\{F_1(1+\cos^2\theta)+F_2(7/3+\cos^2\theta)\right\}\right]}{q(q^2+m_D^2+\delta m_D^2)^2}, 
 \end{align}
 \end{widetext}  
 where the subscript $\text{p}$ stands for perturbative, $m_D^2= g^2T^2\left(\frac{N_c^2}{3}+\frac{N_f}{6}\right)$ is the Debye screening mass in the absence of magnetic field. The quantities $F_1$, $F_2$ are functions of temperature and light quark masses, and have been expressed in detail in Appendix~\ref{AppendixA}. Also,   
 \begin{equation}
     \delta m_D^2=\frac{g^2}{12\pi^2T^2}(q_fB)^2\sum_{l=1}^{\infty}(-1)^{l+1}l^2K_0\left(\frac{m_fl}{T}\right)
 \end{equation}
 is the magnetic field contribution to the Debye screening mass~\cite{Hasan:PRD102'2020,Karmakar:EPJC79'2019}. $K_0$ is modified Bessel function of the second kind.  Once $\text{Im}D^{00}_{\text{p}}$ is evaluated, we obtain $\text{Im}\left[1/\epsilon_Y(q)\right]$ from Eq.~\eqref{imeps}, which eventually is used in Eq.~\eqref{pot} to yield $\text{Im}\,V_Y(q)$:
 \begin{equation}
  \text{Im}\,V_Y(q)=\frac{4}{3}g^2\,\text{Im}D^{00}_{\text{p}}(q_0=0,q).   
 \end{equation}
Similarly, the real part of the potential evaluates to~\cite{Hasan:PRD102'2020}
\begin{align}
  \text{Re}\,V_Y(q)=- \frac{4}{3}g^2\frac{1}{(q^2+m_D^2+\delta m_D^2)},
\end{align}
with $g\equiv g(T,B)$ being the strong coupling.

 \subsection{Non-perturbative (String) part}
As mentioned earlier, the propagator, by definition, being a perturbative entity, does not have a non-perturbative counterpart derived from first principles. However, as a minimal extension of the perturbative scheme [Eq.~\eqref{sde}], a non-perturbative resummed propagator is defined via~\cite{Guo:PRD100'2019}
\begin{equation}\label{imd1}
     D_{R/A}^{L,\text{np}}(Q)=m_G^2\left[q^2-\Pi_{R/A}^{L}(Q)\right]^{-2},
 \end{equation}
where $m_G^2$ is a dimensional constant (of dimension $M^2$) related to the string tension $\sigma$.  For the potential to reduce to the well known vacuum result, $\sigma$ should be related to $m_G^2$ as $\sigma=\frac{2}{3}\alpha_sm_G^2$.  The other difference with Eq.~\eqref{sde} is the power of the terms in the square brackets. Once $D_{R/A}^{L,\text{np}}(Q)$ is evaluated, we use Eq.~\eqref{df} and Eq.~\eqref{imd11} to obtain $\text{Im}D^{00}_{\text{np}}$. The details of the derivation and the final expression is given in Appendix \ref{AppendixB}. The final expression reads
\begin{widetext}
\begin{equation}\label{imdnp}
    \text{Im}D^{00}_{\text{np}}(q_0=0,q)=\frac{-\pi T m_G^2\left[m_D^2-\sum_f\frac{g^2(q_fB)^2}{\pi^2}\left\{F_1(1+\cos^2\theta)+F_2(7/3+\cos^2\theta)\right\}\right]}{q(q^2+m_D^2+\delta m_D^2)^3}.
\end{equation}
\end{widetext}
 
As in the perturbative case, $\text{Im}D^{00}_{\text{np}}$ is then substituted in Eq.~\eqref{imeps} to obtain the non-perturbative contribution of the permitivity, specifically, $\text{Im}\frac{1}{\epsilon_S(q)}$, which finally yields the imaginary part of $V_S(q)$ from Eq.~\eqref{pot}:

\begin{equation}\label{vs}
 \text{Im}\,V_S(q)=\frac{4}{3}g^2\,\text{Im}D^{00}_{\text{np}}(q_0=0,q).   
\end{equation}

For the real part, $\text{Im}D^{00}_{\text{np}}(q_0=0,q)$ is simply replaced by $\text{Re}D^{00}_{\text{np}}(q_0=0,q)$ in Eq.~\eqref{vs}. The result can be expressed as~\cite{Hasan:PRD102'2020} :

\begin{equation}
 \text{Re}\,V_S(q) =- \frac{4}{3}g^2\frac{m_G^2}{(q^2+m_D^2+\delta m_D^2)^2}.  
\end{equation}

Here, the subscript $S$ stands for string. Thus, finally, the complete expression of the in-medium heavy quark potential can be written as:
\begin{widetext}
\begin{align}
    V_Y(q)&=- \frac{4}{3}g^2\left(\frac{1}{q^2+m_D^2+\delta m_D^2}+i\,\frac{\pi T\left[m_D^2-\sum_f\frac{g^2(q_fB)^2}{2\pi^2}\left\{F_1(1+\cos^2\theta)+F_2(7/3+\cos^2\theta)\right\}\right]}{q(q^2+m_D^2+\delta m_D^2)^2}\right)\label{vy}\\[0.3em]
    V_S(q)&=- \frac{4}{3}g^2\left(\frac{m_G^2}{(q^2+m_D^2+\delta m_D^2)^2}+i\,\frac{\pi T m_G^2\left[m_D^2-\sum_f\frac{g^2(q_fB)^2}{\pi^2}\left\{F_1(1+\cos^2\theta)+F_2(7/3+\cos^2\theta)\right\}\right]}{q(q^2+m_D^2+\delta m_D^2)^3}\right)\label{vS},
\end{align}
\end{widetext}
with
\begin{equation}
    V(q)=V_Y(q)+V_S(q).
\end{equation}
 A consistency check is to see if the potential expressed by Eqs.~(\ref{vy}, \ref{vS}) reduces to the vacuum result given in Eq.~\eqref{vac_pot}, under proper conditions. On enforcing $T,B=0$ in Eqs.(\ref{vy}, \ref{vS}), followed by taking a Fourier transform, we obtain
\begin{equation}
V\big|_{T,B=0}(r)\equiv V_0(r)=-\frac{4}{3}\frac{\alpha_s}{r}+ \sigma\, r,   
\end{equation}
which is Eq.~\eqref{vac_pot}. The details are shown in Appendix \ref{AppendixC}. 

\section{SCATTERING RATE}\label{sec4}
The scattering rate is evaluated from the heavy quark self energy, which in turn will be evaluated using the heavy quark potential, as mentioned earlier in Section \ref{sec2}. The Yukawa and string contributions are evaluated separately.

\subsection{Yukawa part}
For the Yukawa part, the usual vector interaction vertex is assumed in the Feynman diagram [Fig.~\ref{figSE}], so that the self energy is written as 
\begin{equation}\label{sigmaY}
    \Sigma_Y(P)=i  \int \frac{d^4 Q}{(2 \pi)^4} V_Y(q) \gamma_\mu S(P-Q) \gamma^\mu.
\end{equation}
Here, we have carried out $g^2D^{\mu\nu}\rightarrow V_Y(q)$ in Eq.~\eqref{self_energy}. Using Eq.\eqref{sigmaY} in Eq.~\eqref{gamma}, the relevant trace would be
\begin{equation}\label{try}
\operatorname{Tr}\left[\gamma_{\mu}(\slashed{P}+M_Q)\gamma^{\mu}(\slashed{P}'+M_Q)\right]=8(M_Q^2+Eq_0-\bm{p}\cdot \bm{q}), 
\end{equation}
where $P_{\mu}'\equiv P_{\mu}-Q_{\mu}$.
The next step is to use the spectral representations of the fermion propagator and the potential. We have for the fermion propagator piece~\cite{Braaten:PRD44'1991}
\begin{align}\label{sff}
		&\frac{1}{K^2-M_Q^2}=\nonumber \\&-\frac{1}{2E'}\int_{0}^{\beta}d\tau'e^{k_0\tau'}\left[(1-n_F(E'))e^{-E'\tau'}-n_F(E')e^{E'\tau'}\right],
	\end{align}	
    where $E'=\sqrt{k^2+M_Q^2}$, $\beta=1/T$, and $n_F$ is the Fermi-Dirac distribution. Similarly, for the potential, we have
\begin{align}\label{sfpot}
	V_Y(q)&=-\int_{0}^{\beta}d\tau\,e^{q_0\tau}\int_{-\infty}^{\infty}d\omega\,\rho_y(q)\left[1+n_B(\omega)\right]e^{-\omega\tau},
\end{align}
with
\begin{equation}
	\rho_Y(q)=-\frac{\text{Im}\,V_Y(q)}{\pi},
\end{equation}
being the spectral function. In the conventional approach, the gluon propagator pieces are expressed via spectral functions.  The spectral representation makes it convenient to carry out the discrete Matsubara frequency summation over $q_0$ using
\begin{subequations}
		\begin{align}
			T\sum_{q_0}e^{q_0(\tau-\tau')}=\delta(\tau-\tau')\label{delta1}\\
			T\sum_{q_0}q_0\,e^{q_0(\tau-\tau')}=\delta'(\tau-\tau')\label{delta2}
		\end{align}
	\end{subequations}
It turns out that the $q_0$ dependent terms in Eq.~\eqref{try} do not give rise to any imaginary part in $\Sigma_Y$, and thus, do not contribute to $\Gamma_Y$~(See~\cite{Bandyopadhyay:PRD105'2022,Dey:PRD109'2024}). Thus, the trace is $8(M_Q^2-\bm{p}\cdot \bm{q})$, which is denoted by $A_Y(q)$.
With all this, the scattering rate expression, after integration over $\tau$, $\tau'$ simplifies to
	\begin{multline}
	\Gamma_Y(E,\bm{v})=\frac{\pi }{4E^2}\int\frac{d^3 q}{(2 \pi)^3}  \int_{-\infty}^{+\infty} d \omega\left[1+n_B(\omega)\right]\rho_Y(q)\\
 A_Y(q)\,\delta(\omega-\bm{v}\cdot\bm{q}),\label{SR}
\end{multline}
where after the frequency sum, the discrete energy $q_0$ is analytically continued to real continuous values via $q_0\rightarrow q_0+i\epsilon$.  
There is a factor $1+n_B(\omega)$ in Eq.~\eqref{SR}. From Eqs. ~\eqref{imd11} and~ \eqref{df}, one can see that the definition of the potential also contains a factor $1+2n_B(\omega)$. The small $\omega$ expansion of these factors are  $T/\omega$ and $2T/\omega$, respectively. Using the formula in Eq. ~\eqref{SR} would then lead to a double counting of the factor $T/\omega$. Hence, we modify the scattering rate formula by removing the aforementioned factor so that it reads
\begin{multline}
	\Gamma_Y(E,\bm{v})=\frac{\pi }{4E^2}\int\frac{d^3 q}{(2 \pi)^3}  \int_{-\infty}^{+\infty} d \omega\rho_Y(q)\\
 A_Y(\omega)\,\delta(\omega-\bm{v}\cdot\bm{q}).\label{SRfinal}
\end{multline}
\subsection{String part}
The string part of the scattering rate is evaluated in a similar fashion. The self energy for the string part is written as
\begin{equation}
\Sigma_S(P)=i  \int \frac{d^4 Q}{(2 \pi)^4} V_S(q)S(P-Q).\label{sigmaS}
\end{equation}
Compared to its Yukawa counterpart [Eq.~\eqref{sigmaY}], the difference is that the Lorentz gamma matrices do not feature here. This is because a scalar interaction vertex is assumed for the string contribution~\cite{Riek:PRC82'2010,Xing:PLB838'2023}. Consequently, the relevant Dirac trace after substituting Eq.~\eqref{sigmaS} in Eq.~\eqref{gamma} is
    \begin{equation}\label{trs}
\operatorname{Tr}\left[(\slashed{P}+M_Q)(\slashed{P}'+M_Q)\right]=4(2M_Q^2-Eq_0+\bm{p}\cdot \bm{q}). 
\end{equation}
The spectral representation of the potential is
\begin{align}\label{sfpotS}
	V_S(q)&=-\int_{0}^{\beta}d\tau\,e^{q_0\tau}\int_{-\infty}^{\infty}d\omega\,\rho_S(q)e^{-\omega\tau},
\end{align}
with
\begin{equation}
	\rho_S(q)=-\frac{\text{Im}\,V_S(q)}{\pi}.
\end{equation}
The spectral function representation of the fermion propagator piece is the same as Eq.~\eqref{sff}. As argued earlier, we have removed the factor $1+n_B(\omega)$ while writing Eq.~\eqref{sfpotS}.
After the Matsubara frequency summation and integration over $\tau$, $\tau'$, we arrive at
\begin{multline}
	\Gamma_S(E,\bm{v})=\frac{\pi }{4E^2}\int\frac{d^3 q}{(2 \pi)^3}  \int_{-\infty}^{+\infty} d \omega\rho_S(q)\\
 A_S(\omega)\,\delta(\omega-\bm{v}\cdot\bm{q}),\label{SRS}
\end{multline}
with  $q_0\rightarrow q_0+i\epsilon$. The $q_0$ dependent term in Eq.~\eqref{trs} does not contribute to $\Gamma_S$, so that, the Dirac trace reads $A_S(q)=4(2M_Q^2+\bm{p}\cdot \bm{q})$. 

\section{MOMENTUM DIFFUSION COEFFICIENTS}\label{sec5}
Once the scattering rate expressions are obtained, the longitudinal and transverse momentum diffusion coefficients can be computed as
\begin{align}
\kappa_{L}&=\int d^3 q \frac{d \Gamma(E,v)}{d^3 q} q_{z}^2\\[0.2em]
\kappa_{T}&=\frac{1}{2}\int d^3 q \frac{d \Gamma(E,v)}{d^3 q} \bm{q_{\perp}}^2
\end{align}
Here, without loss of generality, we assume the magnetic field to be pointing along the $\hat{z}$ direction. In the limit of the heavy quark being static ($p=0=v$), $\omega$ in Eq.~\eqref{SRfinal} is set to 0, owing to the delta function. Using $\Gamma_{Y/S}$, we have
\begin{align}
     (\kappa_L)_{Y/S}&=\frac{1}{16 \pi M_Q^2 }\int_{0}^{\infty}dq\int_{0}^{\pi}d\theta \, q^4 \sin{\theta}\cos^2\theta\, A\,\rho_{Y/S}\label{kL}\\
(\kappa_T)_{Y/S}&=\frac{1}{32 \pi M_Q^2 }\int_{0}^{\infty}dq\int_{0}^{\pi}d\theta \, q^4 \sin^3{\theta}\, A\,\rho_{Y/S}\label{kT},
\end{align}
  The integrals in Eqs.~\eqref{kL}, \eqref{kT} actually turn out to be logarithmically U-V divergent. 
 This is because our calculations are confined to the region of soft gauge boson momentum transfer, and hence, technically incomplete, which manifests in the form of U-V divergence. The integrals, therefore, require a U-V cut-off $q_{\text{max}}$ so that the integrals are
 \begin{align}
     (\kappa_L)_{Y/S}&=\frac{1}{16 \pi M_Q^2 }\int_{0}^{q_{\text{max}}}dq\int_{0}^{\pi}d\theta \, q^4 \sin{\theta}\cos^2\theta\, A\,\rho_{Y/S}\label{kLc}\\
(\kappa_T)_{Y/S}&=\frac{1}{32 \pi M_Q^2 }\int_{0}^{q_{\text{max}}}dq\int_{0}^{\pi}d\theta \, q^4 \sin^3{\theta}\, A\,\rho_{Y/S}\label{kTc}.
\end{align}
Following the prescription of~\cite{Beraudo:NPA831'2009,Bandyopadhyay:2023hiv}, we take the U-V cut-off $q_{\text{max}}$ to be  $3.1Tg(T,B)^{1/3}$, where $g(T,B)$ is the strong coupling that depends on both temperature and magnetic field. In the $B=0$ case, it is shown explicitly that the dependence on this cut-off vanishes once the full range of momentum transfers is taken into account~\cite{Braaten:PRD44'1991}. The same should hold true for weak magnetic fields too, the calculation of which, however, is left for future work. The soft scatterings contribute to $\mathcal{O}(g(T)^2)$ in $\Gamma$, whereas the hard contribution to $\Gamma$ is of $\mathcal{O}(g(T)^4)$. In the case of weak coupling, it can then be inferred that the major contribution to $\Gamma$, and thereby to $\kappa$ of the HQ via elastic scatterings comes from soft gluon exchanges ($\mathcal{O}(gT)$) with the thermal quarks and gluons of the heat bath. However, the relatively rare large-momentum-transfer ($\mathcal{O}(T)$) hard scatterings also play an important role, especially in shaping the logarithmic enhancement that comes from integrating over the full range of momentum transfers. It is also worth mentioning that this U-V cut-off is not necessary if, in the limit of a very strong magnetic field, one uses the Lowest Landau Level (LLL) approximation for writing the HQ propagator. This is because of the presence of the exponential factor $e^{-k_{\perp}^2\big/|q_fB|}$ in the propagator~\cite{Bandyopadhyay:PRD105'2022}.
 
We note that the Dirac traces in the static limit reduce to $A_Y(\omega=0)=A_S(\omega=0)\equiv A=8M_Q^2$. This cancels with the $M_Q^2$ in the denominator of Eqs.~\eqref{kL} and \eqref{kT}, thus rendering the expressions mass-independent.
The only difference between the Yukawa and string contributions of the $\kappa$'s lies in the spectral function $\rho$. As it turns out, the numerical values of $\kappa_L$ and $\kappa_T$ in Eqs.~ \eqref{kL} and \eqref{kT} are exactly similar. Therefore, in the static limit, even in the presence of a weak magnetic field, there is no anisotropy in the momentum diffusion coefficients:
\begin{equation}
    (\kappa_L)_Y=(\kappa_T)_Y, \quad (\kappa_L)_S=(\kappa_T)_S
\end{equation}

\subsection*{Estimation beyond the static limit}

 Eqs.~(\ref{diel}-\ref{pot}) make it clear that the potential approach is valid only in the static limit\cite{Riek:PRC82'2010,Xing:PLB838'2023}. One can, however,  attempt to extrapolate the static limit results to estimate the momentum diffusion coefficients in the dynamic limit where the heavy quark is moving with  certain velocity. Although the accuracy of such an extrapolation remains uncertain beyond the static limit, we dedicate this subsection to provide a qualitative estimation of the non-perturbative effects on the HQ momentum diffusion coefficients in the dynamic limit. The heavy quark velocity $\bm{v}$ and the magnetic field $\bm{B}$, then set the two preferred directions in space, and the interplay between these two directions is used to define the momentum diffusion coefficients. One of the ways of handling the kinematics of the problem is by considering two possible cases; one where the heavy quark velocity is aligned along the direction of the external magnetic field, and the other where the heavy quark velocity lies in a plane perpendicular to the magnetic field. In the following, we provide the expressions for the momentum diffusion coefficients for both cases. The detailed steps are provided in  Appendix~\ref{AppendixD}.

\subsubsection{Case 1: $v\parallel B$}
In this case, we have two momentum diffusion coefficients which are expressed as :
\begin{align}
(\kappa_L)_{Y/S}&=\frac{1}{16 \pi E^2 v}\!\!\!\int\limits_{0}^{q_{\text{max}}}\!\!\! dq 
\!\!\!\int\limits_{-vq}^{vq} \!\!\! d\omega \, q^3\left(\frac{\omega}{vq}\right)^2\!\!\! A_{Y/S}(\omega)\rho_{Y/S}(q) \label{kappaL_bsl_case1}\\
    (\kappa_T)_{Y/S}&=\frac{1}{32 \pi E^2 v} \!\!\!\int\limits_{0}^{q_{\text{max}}}  \!\!\! dq
     \!\!\! \int\limits_{-vq}^{vq}  \!\!\! d\omega\, q^3 \!\!\left(1\!\!-\!\frac{\omega^2}{v^2q^2}\right)\!\! A_{Y/S}(\omega)\rho_{Y/S}(q) \label{kappaT_bsl_case1}
\end{align}

\subsubsection{Case 2: $v\perp B$}
In this case we define three momentum diffusion coefficients considering two anisotropic directions given by $\bm{v}$ and $\bm{B}$, which, after some simplification, can be expressed as :
\begin{align}
&(\kappa_1)_{Y/S}=\frac{1}{16 \pi^2 E^2 v^3}\int\limits_{0}^{q_{\text{max}}}dq\int\limits_{0}^{vq\cos y}d\omega \int\limits_{-\phi'}^{2\pi-\phi'}dy \nn\\
&\times \frac{q\,\omega^2\cos^2{(y+\phi')}}{\cos^3{y}\sqrt{v^2q^2\cos{y}^2-\omega^2}}A_{Y/S}(\omega)\rho_{Y/S}(q), \label{kx} \\
&(\kappa_2)_{Y/S}=\frac{1}{16 \pi^2 E^2 v^3}\int\limits_{0}^{q_{\text{max}}}dq\int\limits_{0}^{vq\cos y}d\omega \int\limits_{-\phi'}^{2\pi-\phi'}dy\nn\\
&\times \frac{q\,\omega^2\sin^2{(y+\phi')}}{\cos^3{y}\sqrt{v^2q^2\cos{y}^2-\omega^2}}A_{Y/S}(\omega)\rho_{Y/S}(q), \label{ky} \\
&(\kappa_3)_{Y/S} =\frac{1}{16 \pi^2 E^2 v^3}\int\limits_{0}^{q_{\text{max}}}dq\int\limits_{0}^{vq\cos y}d\omega \int\limits_{-\phi'}^{2\pi-\phi'}dy \nn\\
&\times \frac{q\,\omega \sqrt{v^2q^2\cos{y}^2-\omega^2}}{\cos^3{y}}A_{Y/S}(\omega)\rho_{Y/S}(q). \label{kz} 
\end{align}

\section{Results}\label{sec6}

\begin{figure*}
	\hspace*{-1cm}
	\begin{minipage}{9cm}
		{\includegraphics[width=0.9\columnwidth]{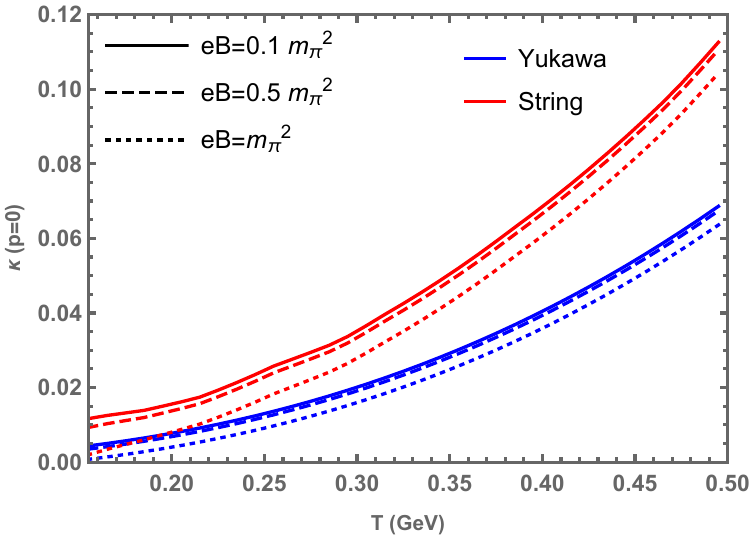}}
	\end{minipage}
\hspace*{0.08cm}
\begin{minipage}{9cm}
		{\includegraphics[width=0.92\columnwidth]{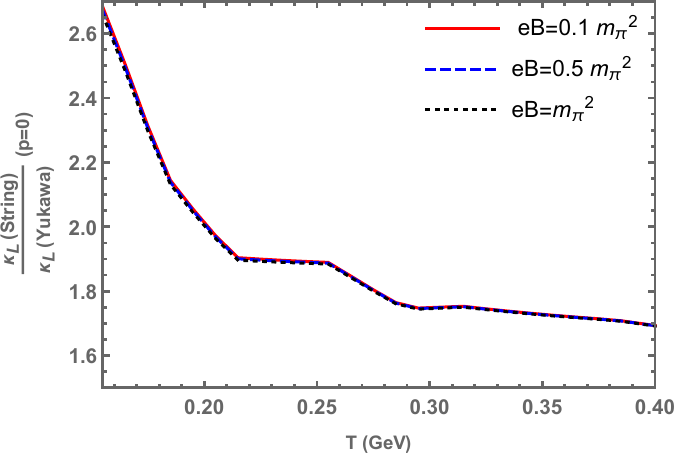}}
	\end{minipage}
   \captionsetup{justification=raggedright, singlelinecheck=false, format=hang, labelsep=period}
    \caption{Left panel: Momentum diffusion coefficient ($\kappa$) in the static limit as a function of temperature. Right Panel: Ratio of string contribution to Yukawa contribution of $\kappa$, as a function of temperature, in the static limit. Both the plots are shown for three different values of the external magnetic field.}
   \label{static limit}
\end{figure*}

The strong coupling $g$ runs with both the magnetic field and temperature~\cite{Ayala:PRD98'2018}
\begin{equation}
    \alpha_s(\Lambda^2,eB)=\frac{g^2}{eB}=\frac{\alpha_s(\Lambda^2)}{1+b_1\alpha_s(\Lambda^2)\ln \frac{\Lambda^2}{\Lambda^2+eB}},
\end{equation}
with
\begin{equation}
    \alpha_s(\Lambda^2)=\frac{1}{b_1\ln\frac{\Lambda^2}{\Lambda^2_{\overline{MS}}}},
\end{equation}
where $\Lambda$ is set at $2\pi T$, $b_1=\frac{11N_c-2N_f}{12\pi}$, $\Lambda_{\overline{MS}}=0.176$ GeV. We have taken $N_c=3$, $N_f=2$. For the finite momentum calculations, we show the results for charm quark for which we take $M_Q=1.28$ GeV. For the zero-momentum results, the mass-dependence cancels out in our calculations, as is expected. As mentioned earlier, the $\kappa$ integrals are logarithmically U-V divergent and hence, require a U-V cut-off, which is taken to be $3.1Tg(T,B)^{1/3}$. The value of the string tension in vacuum, $\sigma_0$, is given by $\sqrt{\sigma_0}=0.465$ GeV~\cite{Riek:PRC82'2010}. There have been string theory studies of the temperature dependence string tension at finite temperature. For instance, in ~\cite{Caristo2022}, a $T$ dependent string tension was defined in the region $0<T<T_c$, given by 
\begin{equation}
    \sigma(T)=\sigma_0\sqrt{1-\frac{\pi T^2}{3\sigma_0}},
\end{equation}
with $\sqrt{\sigma_0}=0.465$ $\text{GeV}$, being the string coupling in vacuum, as mentioned earlier. Beyond $T_c$, the string breaks, and only a spatial string tension can be defined~\cite{Bala:arxiv'2025,Andreev:arxiv'2018}. For our calculations we have used the data of the $T$ dependent spatial string tension presented in \cite{Bala:arxiv'2025}.  

The results of the static limit are shown in Fig.~\ref{static limit}. From the left panel, we see that the momentum diffusion coefficient is a monotonically increasing function of temperature. The Yukawa and the string contributions are also shown separately.  The variation of $\kappa$ with varying magnetic field strength can also be observed in Fig.~\ref{static limit}, where we have shown plots corresponding to three different values of the magnetic field ($eB=0.1~m_{\pi}^2, 0.5~m_{\pi}^2, ~m_{\pi}^2$). For both the Yukawa and string contributions, the magnitude decreases with increasing magnetic field strength. The relative contribution of the Yukawa and string parts to the total $\kappa$ can be quantitatively seen in the right panel of Fig.~\ref{static limit}, where the ratio of the contribution of the string part to the contribution of the Yukawa part is plotted as a function of $T$. At temperatures close to $T_c$, the string part is approximately 2.5 times larger than the Yukawa part. With increase in the temperature, the ratio decreases. This reflects the well-known fact that non-perturbative effects dominate near $T_c$. As mentioned earlier, the longitudinal and transverse contributions are exactly identical. Also, the ratio remains almost unchanged with varying magnetic field strength. This shows that the percentage decrease in magnitude suffered by the string and Yukawa contributions with increasing magnetic field strength are almost identical.

The spatial diffusion coefficient $D_s$ can be defined via the zero-momentum value of $\kappa$, as  mentioned in Eq.~\eqref{ds}:  
\begin{equation}
    D_s=\frac{2T^2}{\kappa(p=0)}
\end{equation}
The $\kappa$ in the above equation is the full $\kappa$ obtained after adding the Yukawa and string contributions, \textit{i.e.}, $\kappa=\kappa_Y+\kappa_S$. 
\begin{figure}
    \begin{center}
     \includegraphics[width=0.89\linewidth]{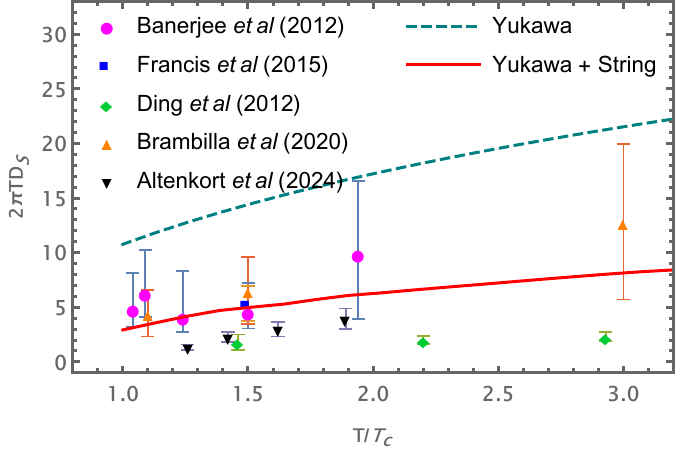}   
    \end{center}
     \captionsetup{justification=raggedright, singlelinecheck=false, format=hang, labelsep=period}
    \caption{Scaled spatial diffusion coefficient as a function of $T$. Several lattice results are also shown for comparison~\cite{Banerjee:PRD85'2012,Francis:PRD92'2015,Ding:PRD86'2012,Brambilla:PRD102'2020,Altenkort:PRL132'2024}. The Altenkort result is HQ mass dependent; their charm quark results are shown in the figure.}
    \label{Dsc}
\end{figure}
Our results for the spatial diffusion coefficient are shown in Fig.~\ref{Dsc}, which can be dubbed as the highlight of the present work. This is an order of magnitude smaller than the perturbative result (shown in the figure by the dashed curve). For comparison, several lattice results have also been shown in the figure. It should be noted that the results presented here are obtained in the presence of a weak background magnetic field, whereas the lattice studies do not take into account any external magnetic field. However, it is also true that the magnetic field considered in our study is weak. It enters into the calculation only as a perturbation about $B=0$ results.\footnote{The weak magnetic field enters into the calculation of the gluon self energy via the fermion propagator. In the presence of a weak magnetic field, the fermion propagator is basically a Taylor series expansion in powers of $q_fB$ around $B=0$.}
  

\begin{figure*}
	\hspace*{-1cm}
	\begin{minipage}{9cm}
		{\includegraphics[width=0.95\columnwidth]{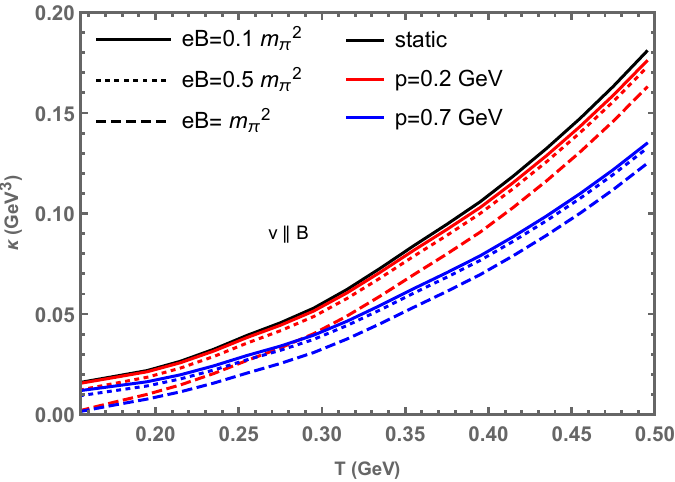}}
	\end{minipage}
\hspace*{0.08cm}
\begin{minipage}{9cm}
		{\includegraphics[width=0.95\columnwidth]{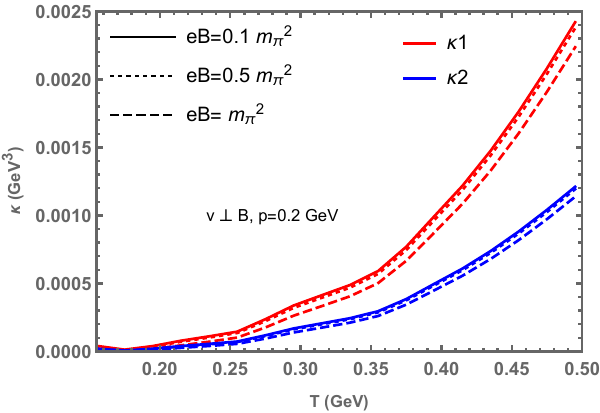}}
	\end{minipage}
    \captionsetup{justification=raggedright, singlelinecheck=false, format=hang, labelsep=period}
    \caption{Left panel: Temperature dependence of momentum diffusion coefficients in the static limit and in the case $\bm{v}\shortparallel \bm{B}$.  Right panel: Temperature dependence of momentum diffusion coefficients in the case $\bm{v}\perp \bm{B}$, with the HQ velocity fixed along $\hat{x}$, \textit{i.e.,} $\phi'=0$. The magnitude of $\kappa_3$ is identical to that of $\kappa_2$. Both the plots are shown for three different values of the external magnetic field.}
    \label{fig:bsl_results}
\end{figure*}

Finally in Fig.~\ref{fig:bsl_results}, we show our estimation of the non-perturbative momentum diffusion coefficients beyond the static limit of the HQ. For this purpose, we have chosen relatively smaller values of the HQ momenta ($p=0.2, 0.7$ GeV), compared to the mass of the charm quark $M_Q$. The left panel showcases the scenario when the HQ is moving along the direction of the magnetic field. Using Eqs.~\eqref{kappaL_bsl_case1} and \eqref{kappaT_bsl_case1}, the magnitudes of $\kappa_L$ and $\kappa_T$ come out to be identical. So, the variation of $\kappa_{L/T}$ with respect to temperature has been denoted by the red ($p=0.2$ GeV) and the blue ($p=0.7$ GeV) curves in the left panel of Fig.~\ref{fig:bsl_results}. When we compare this result with that of the static limit (black solid curve), we discover the qualitative and quantitative similarities between them, although the static limit results dominate the finite-momentum results with $\bm{v}\shortparallel\bm{B}$ for increasing values of $T$. This is more noticeable when we have $p=0.7$ GeV, indicating the decreasing effect of the HQ momentum on the diffusion coefficients.  For each of the momentum value, we plot $\kappa$ for 3 different values of magnetic field strengths to show their dependence on $|\bm{B}|$. Similar to the behavior in the static case, it is seen that the magnitude of $\kappa$ decreases with increasing magnetic field strength.  

In the right panel we display the case $\bm{v}\perp\bm{B}$ with $p=0.2$ GeV, in which the magnitudes for the momentum diffusion coefficients $\kappa_1$ and $\kappa_2(=\kappa_3)$ are much smaller compared to both the $\bm{v}\parallel \bm{B}$ case, and the static case. This shows that for HQs with finite velocity/momentum, the diffusion significantly decreases when the HQ is moving perpendicular to the magnetic field direction. 
The origin of this difference in magnitudes in the trwo cases essentially lies in our choice of the HQ velocity direction.  The trend of coefficient magnitudes with varying magnetic field strengths, however, remains the same as in the $\bm{v}\parallel\bm{B}$ case. The magnitudes decrease with increase in magnetic field strength. The right panel of Fig.~\ref{fig:bsl_results} shows results for the HQ velocity along $\hat{x}$, \textit{i.e.,} for $\phi'=0$. For $\phi'=\pi/2$ (HQ velocity along $\hat{y}$), the black dashed and the blue and red curves get interchanged. This suggests that the momentum transfer between the HQ and the medium happens preferentially along the direction of HQ velocity.

\section{SUMMARY AND CONCLUSIONS}
\label{sec7}
In this work, we have evaluated the momentum and spatial diffusion coefficients ($\kappa$ and $D_s$) of the heavy charm quark propagating through the QGP medium, both for static and dynamic cases, in the presence of a weak background magnetic field. To that end, both perturbative and non-perturbative contributions to $\kappa$ and $D_s$ have been calculated. In order to incorporate the non-perturbative part in our analysis, we take resort to the HQ potential approach, which involves replacing the temporal ($00$) component of the resummed gauge boson propagator with the heavy quark (HQ) in-medium potential. In particular, this replacement is carried out in the calculation of the HQ self energy $\Sigma$ [Eq.~\eqref{self_energy}], whose imaginary part is used, in turn, to calculate the HQ scattering rate $\Gamma$ via Weldon's formula [Eq.~\eqref{gamma}]. By cutting rules, the imaginary part of the resummed HQ self energy can be related to the matrix elements of $t$-channel scattering of the HQ with light thermal partons of the medium. Once $\Gamma$ is evaluated, the momentum diffusion coefficients ($\kappa$) are obtained from Eq.~\eqref{kappa}. In the static HQ case, $D_s$ is then obtained from $\kappa$ via Eq.~\eqref{ds}. Since the potential is in the form of a sum of perturbative and  non-perturbative terms, all the quantities that follow, \textit{viz.} $\Sigma$, $\Gamma$, $\kappa$, are decomposed into perturbative and non-perturbative terms, with the final value given by their addition.

In the static case, the anisotropy direction, in principle, is provided by the magnetic field direction, with respect to which, the longitudinal and transverse momentum diffusion coefficients ($\kappa_L$ and $\kappa_T$) are defined.  We find that in the static case, $\kappa_L=\kappa_T$, \textit{i.e.,} momentum diffusion is isotropic. The string contribution to $\kappa$ is $\sim$5 times larger than the Yukawa contribution at  $T_c$. This ratio decreases with increasing $T$, eventually becoming $<1$ at $T\sim 500$ MeV. In the dynamic scenario, two cases are considered: First, the HQ could be moving parallel to $\bm{B}$, for which, we again obtain $\kappa_L=\kappa_T$.  
For the second case of the HQ velocity lying on a plane perpendicular to $\bm{B}$, we obtain $\kappa_1> \kappa_2\approx\kappa_3$ and the magnitudes in this case to be an order of magnitude smaller than that for the $\bm{v}\parallel \bm{B}$ case. 

In conclusion, non-perturbative contributions to heavy quark dynamics, manifested through the drag and diffusion coefficients, play a significant role in the low temperature region. The implications of this work extend to the study of heavy quark energy loss (both collisional and radiative), nuclear modification factor ($R_{AA}$), and collective flow parameters such as $v_1$ (directed flow )
and $v_2$ (elliptic flow). An intriguing future direction would be to incorporate time dependent magnetic fields while accounting for the medium's electrical conductivity in the analysis.

\section{Acknowledgement}
A.B. is supported by the European Union - NextGenerationEU through grant
No. 760079/23.05.2023, funded by the Romanian ministry of research, innovation and digitalization through Romania’s
National Recovery and Resilience Plan, call no. PNRR-III-C9-2022-I8.  D.D is thankful to the Indian Institute of Technology Bombay for the Institute postdoctoral fellowship. The authors are thankful to Oleg Andreev for his comments. The authors also thank Sayantan Sharma for help with some technicalities.  S. D. acknowledges the SERB Power Fellowship,
SPF/2022/000014 for the support on this work.

\begin{widetext}
	\appendix
	\appendixpage
	\addappheadtotoc
	\begin{appendices}
		\renewcommand \thesubsection{\Alph{section}.\arabic{subsection}}
		\section{Derivation of perturbative resummed propagator [Eq.\eqref{imdp}]}\label{AppendixA}
The starting point is Eq.\eqref{sde}:
\begin{equation}\label{dl}
     D_{R/A}^L(q_0,q)=\left[q^2-\Pi_{R/A}^L(q_0,q)\right]^{-1},
 \end{equation}
 where for brevity, we have dropped the $\text{p}$ superscript. $\Pi^L(q_0,q)\equiv \Pi^{00}(q_0,q)$ is the longitudinal gluon self energy. The general decomposition of the self energy tensor in the presence of a background magnetic field takes the form~\cite{Karmakar:EPJC79'2019}
\begin{equation}
    \Pi^{\mu\nu}(q_0,q)=b(q_0,q)B^{\mu\nu}+c(q_0,q)R^{\mu\nu}+d(q_0,q)Q^{\mu\nu}+a(q_0,q)N^{\mu\nu},
\end{equation} 
where $b$, $c$, $d$, $a$ are Lorentz invariant form factors. $B^{\mu\nu}$, $R^{\mu\nu}$, $Q^{\mu\nu}$, $N^{\mu\nu}$ comprise a set of basis tensors that are mutually orthogonal\footnote{Refer to ~\cite{Karmakar:EPJC79'2019} for exact forms of the tensors}. The longitudinal self-energy simplifies to 
\begin{equation}
    \Pi^{00}(q_0,q)=b(q_0,q)\,\bar{u}^2,
\end{equation}
with $\bar{u}^2=-q^2/Q^2$, where $Q$ is the 4 -momentum of the gluon, and $q=|\bm{q}|$. In the presence of a magnetic field, the form factors split into two parts- magnetic field independent, and magnetic field dependent, so that, we have (suppressing the $q_0$, $q$ dependence)
\begin{equation}
    \Pi^{00}=(b_0+b_2)\bar{u}^2
\end{equation}
$b_2$ is the additional contribution due to the magnetic field. These form factors evaluate to\footnote{There is a relative difference of a minus sign in the expressions with respect to ~\cite{Karmakar:EPJC79'2019}}
\begin{align}
    b_0 \bar{u}^2\equiv\Pi^{00}_{B=0}&=m_D^2\left[\frac{q_0}{2q}\,\ln\frac{q_0+q}{q_0-q}-1\right]\label{b0r}\\
    b_2\bar{u}^2\equiv\Pi^{00}_{B}&=-\delta m_d^2-\sum_f\frac{g^2(q_fB)^2}{\pi^2}\left[F_1(A_0-A_2)+F_2(5A_0/3-A_2)\right],\label{b2r}
\end{align}
where, 
\begin{align}
    A_0&=\frac{q_0}{2q}\,\ln \frac{q_0+q}{q_0-q}\\[0.3em]
    A_2&=\frac{q_0^2}{2 q^2}\left(1-\frac{3q_3^2}{q^2}\right)\left(1-\frac{q_0}{2 q} \log \frac{q_0+q}{q_0-q}\right) +\frac{1}{2}\left(1-\frac{q_3^2}{q^2}\right) \frac{q_0}{2 q} \log \frac{q_0+q}{q_0-q}
\end{align}
$F_1$, $F_2$ are $T$ dependent functions:
\begin{equation}
    F_1=g_k+\frac{\pi m_f-4T}{32 m_f^2T}\,,\quad F_2=f_k+\frac{8T-\pi m_f}{128 m_f^2T},
\end{equation}
with\footnote{$K_0$, $K_1$ are Bessel functions of the second kind.}
\begin{equation}
    g_k=\sum_{l=1}^{\infty}(-1)^{l+1}\frac{l}{4m_fT}K_1\left(\frac{m_fl}{T}\right)\,,\quad     f_k=-\sum_{l=1}^{\infty}(-1)^{l+1}\frac{l^2}{16T^2}K_2\left(\frac{m_fl}{T}\right).
\end{equation}
The imaginary parts come from from the $A_n$ functions by continuing $q_0$ to $q_0+i\epsilon$ (retarded) or to ${\bl q_0-i\epsilon}$ (advanced). Thus, up to $\mathcal{O}(q_0/q)$, we have
\begin{align}
   b_0^R\,\bar{u}^2&=-\frac{i\pi}{2}m_D^2\frac{q_0}{q}-m_D^2 \\
   b_2^R\,\bar{u}^2&=-\delta m_D^2-\sum_f\frac{g^2(q_fB)^2}{\pi^2}\left[F_1(A_0^R-A_2^R)+F_2(5A_0^R/3-A_2^R)\right],
\end{align}
with 
\begin{equation}\label{ar}
    A_0^R=\frac{-i\pi q_0}{2q}\,,\quad A_2^R=\frac{1}{2}\left(1-\frac{q_3^2}{q^2}\right)\frac{q_0}{2q}\left(\frac{-i\pi}{2}\right).
\end{equation}
Using Eqs.[\ref{ar}, \ref{b0r}, \ref{b2r}] in Eq.\eqref{dl}, we obtain for the retarded propagator
\begin{equation}
    (D^L)_R=\left[q^2+m_D^2+\frac{i\pi}{2}m_D^2\frac{q_0}{q}+\delta m_D^2+\sum_f\frac{g^2(q_fB)^2}{\pi^2}\left\{F_1(A_0^R-A_2^R)+F_2(5A_0^R/3-A_2^R)\right\}      \right]^{-1}\equiv X
\end{equation}
Similarly, for the advanced propagator, we have
\begin{equation}
    (D^L)_A=\left[q^2+m_D^2-\frac{i\pi}{2}m_D^2\frac{q_0}{q}+\delta m_D^2+\sum_f\frac{g^2(q_fB)^2}{\pi^2}\left\{F_1(A_0^A-A_2^A)+F_2(5A_0^A/3-A_2^A)\right\}      \right]^{-1}\equiv Y,
\end{equation}
where, 
\begin{equation}
    A_0^R=-A_0^A\,,\quad A_2^R=-A_2^A
\end{equation}
Further, we have
\begin{equation}
A_0^R-A_2^R=\frac{-i\pi q_0}{4q}(1+\cos^2\theta)\,,\quad \frac{5}{3}A_0^R-A_2^R=\frac{-i\pi q_0}{4q}(7/3+\cos^2\theta),    
\end{equation}
with $\cos^2\theta=q_3^2/q^2$. Let $q^2+m_D^2+\delta m_D^2=a$. Then, 
\begin{multline}
    \frac{1}{X}\cdot \frac{1}{Y}=a^2-\frac{i\pi}{2}m_D^2\frac{q_0}{q}a+\frac{i\pi q_0}{4q}a\sum_f\frac{g^2(q_fB)^2}{\pi^2}\left\{F_1(1+\cos^2\theta)+F_2(7/3+\cos^2\theta)\right\}+\frac{i\pi}{2}m_D^2\frac{q_0}{q}a
    +\frac{\pi^2}{4}m_D^2\left(\frac{q_0}{q}\right)^2\\
    -\frac{\pi^2}{4}m_D^2\left(\frac{q_0}{q}\right)^2\sum_f\frac{g^2(q_fB)^2}{\pi^2}\left\{F_1(1+\cos^2\theta)+F_2(7/3+\cos^2\theta)\right\}
    -\frac{i\pi q_0}{4q}a\sum_f\frac{g^2(q_fB)^2}{\pi^2}\left\{F_1(1+\cos^2\theta)+F_2(7/3+\cos^2\theta)\right\}\\+\frac{\pi^2}{4}\left(\frac{q_0}{q}\right)^2\left[\sum_f\frac{g^2(q_fB)^2}{\pi^2}\left\{F_1(1+\cos^2\theta)+F_2(7/3+\cos^2\theta)\right\}\right]^2
\end{multline}
Up to $\mathcal{O}\left(q_0/q\right)$
\begin{equation}\label{prod}
    \frac{1}{X}\cdot \frac{1}{Y}=a^2=(q^2+m_D^2+\delta m_D^2)^2
\end{equation}
Also, \begin{equation}\label{diff}
 \frac{1}{X}-\frac{1}{Y}=i\pi m_D^2\frac{q_0}{q}-\frac{i\pi q_0}{2q} \sum_f\frac{g^2(q_fB)^2}{\pi^2}\left\{F_1(1+\cos^2\theta)+F_2(7/3+\cos^2\theta)\right\}  
\end{equation}
We note that
\begin{equation}\label{xy}
    X-Y=-\left(\frac{1}{X}-\frac{1}{Y}\right)\frac{1}{\frac{1}{X}\cdot \frac{1}{Y}}.
\end{equation}
Then, using Eq.\eqref{prod} and Eq.\eqref{diff}, we have
\begin{equation}
    X-Y\equiv  (D^L)_R-(D^L)_A=\frac{-i\pi \frac{q_0}{q}\left[m_D^2-\sum_f\frac{g^2(q_fB)^2}{\pi^2}\left\{F_1(1+\cos^2\theta)+F_2(7/3+\cos^2\theta)\right\}\right]}{(q^2+m_D^2+\delta m_D^2)^2}
\end{equation}
From Eqs.[\ref{df}, \ref{df_exp}], for $q_0/T\ll 1$, we have for the Feynman propagator
\begin{equation}
    \frac{1}{2}D^L_F=\frac{T}{q_0}\left[(D^L)_R-(D^L)_A\right].
\end{equation}
Hence, from Eq.\eqref{imd11}
\begin{equation}
    \text{Im}D^L(q_0=0,q)=\frac{-\pi T \left[m_D^2-\sum_f\frac{g^2(q_fB)^2}{\pi^2}\left\{F_1(1+\cos^2\theta)+F_2(7/3+\cos^2\theta)\right\}\right]}{(q^2+m_D^2+\delta m_D^2)^2}.
\end{equation}
This is Eq.\eqref{imdp} in the text.

\section{Derivation of non-perturbative resummed propagator [Eq.\eqref{imdnp}]}\label{AppendixB}
 The starting point is Eq.\eqref{imd1} (suppressing the np superscript):
 \begin{equation}
     D_{R/A}^{L}(Q)=m_G^2\left[q^2-\Pi_{R/A}^{L}(Q)\right]^{-2},
 \end{equation}
where $m_G^2$ is independent of $q_0,q$. Proceeding as earlier, we obtain for the retarded propagator and advanced propagators:
\begin{equation}
    (D^L)_R=m_G^2\left[q^2+m_D^2+\frac{i\pi}{2}m_D^2\frac{q_0}{q}+\delta m_D^2+\sum_f\frac{g^2(q_fB)^2}{\pi^2}\left\{F_1(A_0^R-A_2^R)+F_2(5A_0^R/3-A_2^R)\right\}      \right]^{-2}\equiv X
\end{equation}
\begin{equation}
    (D^L)_A=m_G^2\left[q^2+m_D^2-\frac{i\pi}{2}m_D^2\frac{q_0}{q}+\delta m_D^2+\sum_f\frac{g^2(q_fB)^2}{\pi^2}\left\{F_1(A_0^A-A_2^A)+F_2(5A_0^A/3-A_2^A)\right\}      \right]^{-2}\equiv Y
    \end{equation}

Thus
\begin{align}
    \text{Num}\left(\frac{1}{X}\right)&=\left[q^2+m_D^2+\delta m_D^2+\frac{i\pi}{2}m_D^2\frac{q_0}{q}-\frac{i\pi q_0}{4q}\sum_f\frac{g^2(q_fB)^2}{\pi^2}\left\{F_1(1+\cos^2\theta)+F_2(7/3+\cos^2\theta)\right\}\right]^2\\
     \text{Num}\left(\frac{1}{Y}\right)&=\left[q^2+m_D^2+\delta m_D^2-\frac{i\pi}{2}m_D^2\frac{q_0}{q}+\frac{i\pi q_0}{4q}\sum_f\frac{g^2(q_fB)^2}{\pi^2}\left\{F_1(1+\cos^2\theta)+F_2(7/3+\cos^2\theta)\right\}\right]^2
\end{align}
After a bit of straightforward simplification, we have, up to $\mathcal{O}(q_0/q)$,
\begin{equation}\label{b6}
  \frac{1}{X}\cdot \frac{1}{Y}=\frac{(q^2+m_D^2+\delta m_D^2)^4}{m_G^4}.
\end{equation}
To extract $\text{Num}\left(\frac{1}{X}-\frac{1}{Y}\right)$ from $\text{Num}\left(\frac{1}{X}\right)$, $\text{Num}\left(\frac{1}{Y}\right)$ , we use the relation $(x+iy)^2-(x-iy)^2=4xy$
\begin{align}
  \text{Num} \left( \frac{1}{X}-\frac{1}{Y}\right)&=4i\left(q^2+m_D^2+\delta m_D^2\right)\frac{\pi q_0}{2q}\left[m_D^2-\sum_f\frac{g^2(q_fB)^2}{\pi^2}\left\{F_1(1+\cos^2\theta)+F_2(7/3+\cos^2\theta)\right\}\right]\label{nn}\\
  \text{Den}\left(\frac{1}{X}-\frac{1}{Y}\right)&=m_G^2\label{nd}
\end{align}
 We use Eqs.[\ref{xy}, \ref{b6}, \ref{nn}, \ref{nd}] in Eq.\eqref{xy} to obtain 
 \begin{equation}
   X-Y  \equiv  (D^L)_R-(D^L)_A=\frac{-i\pi\frac{q_0}{q}m_G^2\left[m_D^2+\sum_f\frac{g^2(q_fB)^2}{\pi^2}\left\{F_1(1+\cos^2\theta)+F_2(7/3+\cos^2\theta)\right\}\right]}{(q^2+m_D^2+\delta m_D^2)^3}.
 \end{equation}
The Feynman propagator is then evaluated via
\begin{equation}
\frac{1}{2}D^L_F=\frac{T}{q_0}\left[(D^L)_R-(D^L)_A\right].    
\end{equation}
Finally, from Eq.\eqref{imd11}, we get, for the non-perturbative resummed propagator
\begin{equation}
    \text{Im}D^{00}_{\text{np}}(q_0=0,q)=\frac{-\pi T m_G^2\left[m_D^2-\sum_f\frac{g^2(q_fB)^2}{\pi^2}\left\{F_1(1+\cos^2\theta)+F_2(7/3+\cos^2\theta)\right\}\right]}{q(q^2+m_D^2+\delta m_D^2)^3}.
\end{equation}
This is Eq.\eqref{imdnp} in the text.

\section{Reduction of the in-medium HQ potential to the vacuum result}
\label{AppendixC}
From Eq.\eqref{vy} and Eq.\eqref{vS}, we have
\begin{align}
    \text{Re}\,V(q)=\text{Re}\,V_Y(q)+\text{Re}\,V_S(q)=-\frac{4}{3}4\pi\alpha_s\left[\frac{1}{q^2+m_D^2+\delta m_D^2}+\frac{m_G^2}{(q^2+m_D^2+\delta m_D^2)^2}\right],
\end{align}
 with $\alpha_s=g^2/(4\pi)$. Similarly,
\begin{multline}
  \text{Im}\,V(q)= -\frac{4}{3}4\pi\alpha_s\pi T\Bigg(\frac{m_D^2-\sum_f\frac{g^2(q_fB)^2}{2\pi^2}\left\{F_1(1+\cos^2\theta)+F_2(7/3+\cos^2\theta)\right\}}{q(q^2+m_D^2+\delta m_D^2)^2}+\\\frac{ m_G^2\left[m_D^2-\sum_f\frac{g^2(q_fB)^2}{\pi^2}\left\{F_1(1+\cos^2\theta)+F_2(7/3+\cos^2\theta)\right\}\right]}{q(q^2+m_D^2+\delta m_D^2)^3}\Bigg) 
\end{multline}

Setting $T,B=0$ (vacuum), we get
\begin{align}
    \text{Re}\,V_0(q)= -\frac{4}{3}4\pi\alpha_s\left(\frac{1}{q^2}+\frac{m_G^2}{q^4}\right)\,,\qquad \text{Im}\,V_0(q)=0.
\end{align}
The vacuum potential is purely real. Thus, we have in vacuum
\begin{equation}
   V(q)\bigg|_{T=0}= -\frac{4}{3}\frac{4\pi\alpha_s}{q^2}-\frac{8\pi\sigma}{q^4}\equiv  V_0^{\text{p}}(q)+V_0^{\text{np}}(q)\equiv V_0(q),
\end{equation}
where, we have used $\sigma=2\alpha_s m_G^2/3$. Taking the Fourier transform of the respective terms, we arrive at
\begin{equation}
    V_0^{\text{p}}(q)\xrightarrow[F.T.]{} -\frac{4}{3}\frac{\alpha_s}{r}\equiv V_0^\text{p}(r).
\end{equation}
\begin{equation}
     V_0^{\text{np}}(q)\xrightarrow[F.T.]{}\sigma\, r\equiv V_0^{\text{np}}(r).
\end{equation}
 Thus, the vacuum potential in the coordinate space comes out to be
 \begin{equation}
     V_0(r)=-\frac{4}{3}\frac{\alpha_s}{r}+\sigma\, r.
 \end{equation}

 \section{Evaluation of momentum diffusion coefficients beyond the static limit}
 \label{AppendixD}
\subsection{Case 1: $v\parallel B$}
In the first case, we consider the HQ velocity to be along the direction of magnetic field, so that, effectively, we still have a single preferred direction in space, $\hat{z}$. Then, we have
\begin{align}
		(\kappa_L)_{Y/S}=\frac{\pi }{4E^2}\sum_{i=1}^4\int\frac{d^3 q}{(2 \pi)^3} q_L^2 \int_{-\infty}^{+\infty} d \omega \,\rho_{Y/S}(q) A^{\parallel}_{Y/S}(\omega)\,\delta(\omega-\bm{v}\cdot\bm{q}).\label{kappa2L}
  \end{align}
  \begin{align}
     (\kappa_L)_{Y/S}=\frac{\pi }{8E^2}\sum_{i=1}^4\int\frac{d^3 q}{(2 \pi)^3} q_T^2 \int_{-\infty}^{+\infty} d \omega \,\rho_{Y/S}(q) A^{\parallel}_{Y/S}(\omega)\,\delta(\omega-\bm{v}\cdot\bm{q}),
  \end{align}
 where $q_L^2=q^2\cos^2\theta$, $q_T^2=q^2-q_L^2=q^2(1-\cos^2\theta)$. Because the HQ velocity points along $\bm{B}$ (hence, along $\hat{z}$), we have $\bm{v}\cdot\bm{q}=vq \cos \theta\equiv vq\eta$, where, $\theta$ is both the angle between $\bm{q}$ and $\bm{v}$, as well as the polar angle of integration in the above equations. The delta function is then used to integrate over $\eta$ with $d^3q=2\pi q^2dqd\eta$, which sets $\omega=vq\eta$. Since $-1\leq \eta\leq 1$, $-vq\leq \omega\leq vq$. This finally leads to Eqs.~\eqref{kappaL_bsl_case1} and \eqref{kappaT_bsl_case1}.

\subsection{Case 2: $v\perp B$}
The HQ velocity now lies in the $x$-$y$ plane, with the magnetic field pointing in the $\hat{z}$ direction, as earlier. Since $\bm{v}$ is no longer oriented along the $z$ axis, $\bm{v}\cdot\bm{q}$ is not trivial. The direction of HQ velocity in the $x$-$y$ plane can be specified by the azimuthal angle $\phi'$ (The polar angle $\theta'=0$, as the velocity vector is in the $x$-$y$ plane). The vector $\bm{q}$ is specified by $q$, $\theta$ and $\phi$ (our integration variables), where $\theta$ is the polar angle, and $\phi$ is the azimuthal angle. Then,
\begin{equation}
	\bm{v}\cdot\bm{q}=vq\sin\theta\cos(\phi-\phi'),
\end{equation}
where $v=|\bm{v}|$, $q=|\bm{q}|$. Then, using Eq.~\ref{SRfinal}, the interaction rate becomes
\begin{align}
			\Gamma_{Y/S}(E,\bm{v})=\frac{\pi}{4E^2(2\pi)^3}\int dq\,q^2\int_0^\pi d\theta\sin\theta\int_0^{2\pi}d\phi \int_{-\infty}^{+\infty} d \omega\,
			\rho_{Y/S}( q) A^{\perp}_{Y/S}(\omega)\,\delta[\omega-vq\sin\theta\cos(\phi-\phi')]\label{gamma_perp}
	\end{align}	
It turns out that $A_{\parallel}=A_{\perp}$, which we denote simply by $A$. We introduce a variable $y=\phi-\phi'$. $\Gamma$ then simplifies to
\begin{align}
	\Gamma_{Y/S}(E,\bm{v})=\frac{1}{32E^2v\pi^2}\int dq\,q\int_0^\pi d\theta \int_{-\phi'}^{2\pi-\phi'}dy \sin\theta\,\delta\left(\sin\theta-\frac{\omega}{vq\cos y}\right)\frac{1}{\cos{y}}\int_{-\infty}^{+\infty} d \omega A_{Y/S}(\omega)\rho_{Y/S}(q).
\end{align} 
For the case $\bm{v}\perp \bm{B}$, three momentum diffusion coefficients are defined conventionally: 
\begin{align}
(\kappa_{1})_{Y/S}&=\int d^3 q \frac{d \Gamma_{Y/S}(E,\bm{v})}{d^3 q} q_{x}^2.\label{k1}\\[0.2em]
(\kappa_{2})_{Y/S}&=\int d^3 q \frac{d \Gamma_{Y/S}(E,\bm{v})}{d^3 q} q_y^2.\label{k2}\\[0.2em]
(\kappa_{3})_{Y/S}&=\int d^3 q \frac{d \Gamma_{Y/S}(E,\bm{v})}{d^3 q} q_z^2.\label{k3}
\end{align} 
Here, $q_x^2=q^2\sin^2\theta\cos^2(\phi'+y)$, $q_y^2=q^2\sin^2\theta\sin^2(\phi'+y)$, $q_z^2=q^2\cos^2\theta$.  The next step is to evaluate the expressions of $\kappa_{1,2,3}$ by using $\Gamma(E,\bm{v})$ from Eq.\eqref{gamma_perp}. To do the $\theta$ integrations, we use the following results
\begin{equation}
    \int_0^\pi d\theta \sin^n\theta\,\delta\left(\sin\theta-c\right)=\frac{2c^n}{\sqrt{1-c^2}}\Theta(c)\Theta(1-c)
    \end{equation}
    \begin{equation}
        \int_0^\pi d\theta \sin\theta\cos^2\theta\,\delta\left(\sin\theta-c\right)=2c\sqrt{1-c^2}\,\Theta(c)\Theta(1-c),
    \end{equation}
where $c\in \mathbb{R}$, and $n$ is an integer. The $\Theta$ function sets 
\begin{align}
	&0\leq \frac{\omega}{v q\cos y}\leq 1, \nn\\ 
	&0\leq \omega\leq vq\cos y
\end{align}
Thus, the final expressions come out to be Eqs.~\eqref{kx}, \eqref{ky} and \eqref{kz}.

\end{appendices}
\end{widetext}

\bibliography{ref}{}
\end{document}